\documentstyle[12pt,epsf]{article}

\begin{document}

\renewcommand{\thepage}{}

\begin{titlepage}


\title{
\hfill
\parbox{5cm}{\normalsize ICRR-Report-369-96-20\\
hep-ph/yymmdd}\\
\vspace{5ex}
CP and T violation test in neutrino oscillation
\vspace{5ex}}
\author{
Jiro Arafune\thanks{e-mail address:
 {\tt arafune@icrhp3.icrr.u-tokyo.ac.jp}}
and
Joe Sato\thanks{e-mail address:
 {\tt joe@icrhp3.icrr.u-tokyo.ac.jp}}\\
 {\it Institute for Cosmic Ray Research,
   University of Tokyo, Midori-cho,}\\
   {\it   Tanashi, Tokyo 188, Japan}}
\date{\today}

\maketitle

\begin{abstract}
We examine how large violation of CP and T is allowed
in long base line neutrino experiments.
When we attribute only the atmospheric neutrino anomaly to
neutrino oscillation we may have large CP violation
effect.
When we attribute both the atmospheric neutrino anomaly
and the solar neutrino deficit to
neutrino oscillation we may have a sizable T violation
effect proportional to the ratio of two mass differences;
it is difficult to see CP violation since we can't ignore
the matter effect. We give a simple expression for T violation
in the presence of matter.
\end{abstract}

\end{titlepage}

\newpage
\renewcommand{\thepage}{\arabic{page}}

\section{Introduction}

The CP or T violation is a fundamental and important problem of the
particle physics and cosmology.
The CP study of the lepton sector, though it has been less 
examined than that of quark sector, is
indispensable, since the neutrinos are allowed to have 
masses and complex mixing angles in the electroweak theory.

The neutrino oscillation search is a powerful experiment which can
examine masses and mixing angles of the neutrinos.
In fact the several underground experiments have shown
lack of the solar
neutrinos\cite{Ga1,Ga2,Kam,Cl}
and anomaly in the atmospheric
neutrinos\cite{AtmKam,IMB,NUSEX,SOUDAN2,Frejus},
implying that there may occur the neutrino oscillation. The
atmospheric neutrino anomaly suggests mass difference
around $10^{-3} \sim 10^{-2}$
eV$^2$\cite{Fogli1,FLM,Yasuda}, which encourages us to make
long base line neutrino experiments. Recently such
experiments are planned and will be operated in the
near future\cite{KEKKam,Ferm}. It seems necessary for 
us to examine whether there is a chance to observe 
not only the neutrino oscillation but also the CP or
T violation by
long base line experiments. In this short paper we study
such possibilities
taking account of the atmospheric neutrino experiments and
also considering the solar neutrino experiments and others.

\section{Formulation of
CP and T violation in neutrino oscillation}

\subsection{Brief review}

We briefly review CP and T
violation in vacuum oscillation\cite{yana,BP,Pakvasa}
to clarify our notation.

Let's denote the mass eigenstates
of 3 generations of neutrinos by $\nu_m = (\nu_1,\nu_2,\nu_3)$
with mass eigenvalues\footnote{
We assume $m_1 < m_2 < m_3$ in vacuum.} $(m_1, m_2, m_3)$ and
the weak eigenstates by $\nu_w = (\nu_e,\nu_\mu,\nu_\tau)$
corresponding to electron, $\mu$ and $\tau$, respectively.
They are connected
by a unitary transformation:

\begin{eqnarray}
\nu_w = U \nu_m,
  \label{eq:numixing}
\end{eqnarray}
where $U$ is a unitary ($3 \times 3$) matrix
similar to the CKM matrix for quarks.
We will use the parametrisation for $U$
by Chau and Keung\cite{CK,KP,Toshev},
\begin{eqnarray}
U &=& \pmatrix{1&0&0\cr 0&c_\psi&s_\psi\cr 0&-s_\psi&c_\psi}
\pmatrix{1&0&0\cr 0&1&0\cr 0&0&e^{i \delta}}
\pmatrix{c_\phi&0&s_\phi\cr 0&1&0\cr -s_\phi&0&c_\phi}
\pmatrix{c_\omega&s_\omega&0\cr -s_\omega&c_\omega&0\cr 0&0&1}
\label{eq:mixingmatrix}\\
&=&{\rm exp}(i \psi \lambda_7) \Gamma\ {\rm exp}(i \phi \lambda_5)
{\rm exp}(i \omega \lambda_2),
\end{eqnarray}
where the $\lambda$'s are the Gell-Mann matrices.

The evolution equation for the weak eigenstate is given by
\begin{eqnarray}
i{ {\rm d} \over {\rm d}x} \nu_w &=& 
- U {\rm diag} (p_1, p_2, p_3)  U^\dagger \nu_w
\nonumber \\
&\simeq& \{- p_1 + {1 \over 2E} U {\rm diag}
(0, \delta m_{21}^2, \delta m_{31}^2) U^\dagger\} \nu_w
\nonumber \\
&\sim& {1 \over 2E} U {\rm diag}
(0, \delta m_{21}^2, \delta m_{31}^2) U^\dagger \nu_w,
\label{eq:evoleq}
\end{eqnarray}
where $p_i$'s are the momenta,
$E$ is the energy and
$\delta m_{ij}^2 = m_i^2 - m_j^2$.
A term proportional to a unit matrix like
$p_1$ in eq.\ref{eq:evoleq} is dropped
because
it is irrelevant to the transition
probability.

The solution for the equation is
\begin{eqnarray}
\nu_w(x) =  U \exp( -i
{x \over 2E}{\rm diag}
(0, \delta m_{21}^2, \delta m_{31}^2) ) U^\dagger \nu_w(0).
\end{eqnarray}

The transition probability of $\nu_\alpha \rightarrow
\nu_\beta (\alpha, \beta = e, \mu, \tau)$ at distance $L$ 
is given by
\begin{eqnarray}
P(\nu_\alpha \rightarrow \nu_\beta; E, L) &=&
|\sum_{i,j} U_{\beta i} (e^{ -i
{L \over 2E}{\rm diag}
(0, \delta m_{21}^2, \delta m_{31}^2 })_{ij} U^*_{\alpha j} |^2\\
&=&\sum_{i,j} U_{\beta i}U^*_{\beta j}U^*_{\alpha i}U_{\alpha j}
\exp \{-i \delta m^2_{ij}(L/2E) \}.
\end{eqnarray}

The T violation gives the difference
between the transition probability of
$\nu_\alpha \rightarrow \nu_\beta$ and that of
$\nu_\beta \rightarrow \nu_\alpha$\cite{BWP}:
\begin{eqnarray}
&&P(\nu_\alpha \rightarrow \nu_\beta; E, L) -
P(\nu_\beta \rightarrow \nu_\alpha; E, L)
\nonumber \\
&=&-4 ({\rm Im}U_{\beta 1}U^*_{\beta 2}U^*_{\alpha 1}U_{\alpha 2})
(\sin\Delta_{21}+\sin\Delta_{32}+\sin\Delta_{13})
\label{eq:BWP}
\\
&\equiv& 4 J f,
\label{eq:jf}
\end{eqnarray}
where 
\begin{eqnarray}
\Delta_{ij}&\equiv&\delta m_{ij}^2 {L\over 2E}
=2.54 {(\delta m_{ij}^2/10^{-2}{\rm eV}^2)
\over (E/{\rm GeV})} (L/100{\rm km}),\\
J &\equiv& -{\rm Im}U_{\beta 1}U^*_{\beta 2}U^*_{\alpha 1}U_{\alpha 2},
\label{eq:Jdef}\\
f &\equiv&
(\sin\Delta_{21}+\sin\Delta_{32}+\sin\Delta_{13})
\label{eq:fdef}
\\
&=&-4\sin{\Delta_{21}\over 2}\sin{\Delta_{32}\over 2}
\sin{\Delta_{13}\over 2}.
\label{eq:fv}
\end{eqnarray}

The unitarity of $U$ gives
\begin{eqnarray}
J=\pm \sin\omega \cos\omega \sin\psi \cos\psi
\sin\phi \cos^2\phi \sin\delta
\label{eq:J}
\end{eqnarray}
with the sign $+$ ($-$) for $\alpha, \beta$ in
cyclic (anti-cyclic) order($+$ for $(\alpha,\beta)
=(e,\mu), (\mu,\tau)$ or $(\tau,e)$). In the following
we assume the cyclic order for ($\alpha$ ,
$\beta$) for simplicity.

There are bounds for $J$ and $f$,

\begin{equation}
|J|\le {1\over 6\sqrt{3}},
\label{eq:Jmax}
\end{equation}
where the equality holds for
$|\sin\omega|=1/\sqrt{2},|\sin\psi|=1/\sqrt{2},
|\sin\phi|=1/\sqrt{3}$ and $|\sin\delta|=1$, and\cite{Cabibbo}

\begin{eqnarray}
|f|\le {3 \sqrt 3 \over 2},
\label{eq:fmax}
\end{eqnarray}
where the equality holds for
$\Delta_{21} \equiv \Delta_{32} \equiv 2\pi/3 $ (mod $2\pi$).

In the vacuum the CPT theorem gives the relation between
the transition probability of
anti-neutrino and that of neutrino,
\begin{eqnarray}
P(\bar\nu_\alpha \rightarrow \bar\nu_\beta; E, L) =
P(\nu_\beta \rightarrow \nu_\alpha; E, L),
\end{eqnarray}
which relates CP violation to T violation:
\begin{eqnarray}
&&P(\nu_\alpha \rightarrow \nu_\beta; E, L) -
P(\bar\nu_\alpha \rightarrow \bar\nu_\beta; E, L)
\nonumber \\
&=&P(\nu_\alpha \rightarrow \nu_\beta; E, L) -
P(\nu_\beta \rightarrow \nu_\alpha; E, L).
\label{eq:cpt}
\end{eqnarray}

\subsection{CP and T violation with disparate
mass differences}

Let's consider how large the T(CP) violation
can be in the ``disparate'' mass difference case\footnote{
Hereafter we denote the larger mass difference by
$\delta m^2_{31}$ and the smaller one by $\delta m^2_{21}$
in the case that the mass differences have a
large ratio.},
say $\epsilon \equiv {\delta m^2_{21} \over \delta m^2_{31}}
\ll 1$. In this case
the following two situations are interesting\cite{BWP},
since in the case $\Delta_{31} \ll 1$ we
have too small $f (\sim
O(\epsilon \Delta^3_{31})$ due to eq.\ref{eq:fv})
to observe the T(CP) violation effect:

\noindent
Situation 1. $\Delta_{31} \sim O(1)$.

Because $|\epsilon \Delta_{31}| \ll 1$ in this case,
the oscillatory part $f$
becomes $O(\epsilon)$:
\begin{eqnarray}
f(\Delta_{31},\epsilon)
&=&\sin\Delta_{21}+\sin\Delta_{32}+\sin\Delta_{13}
\nonumber\\
&=&\sin(\epsilon\Delta_{31})+\sin\{(1-\epsilon)\Delta_{31}\}
-\sin\Delta_{31}
\label{eq:f}\\
&=& \epsilon \Delta_{31}(1-\cos\Delta_{31})
+O(\epsilon^2 \Delta_{31}^2).
\label{eq:fapp}
\end{eqnarray}

\begin{figure}
\unitlength=1cm
\begin{picture}(16,6)
\unitlength=1mm
\put(25,80){$f(\Delta_{31},\epsilon = 0.03)$}
\put(150,5){$\Delta_{31}$}
\centerline{
\epsfxsize=13cm
\epsfbox{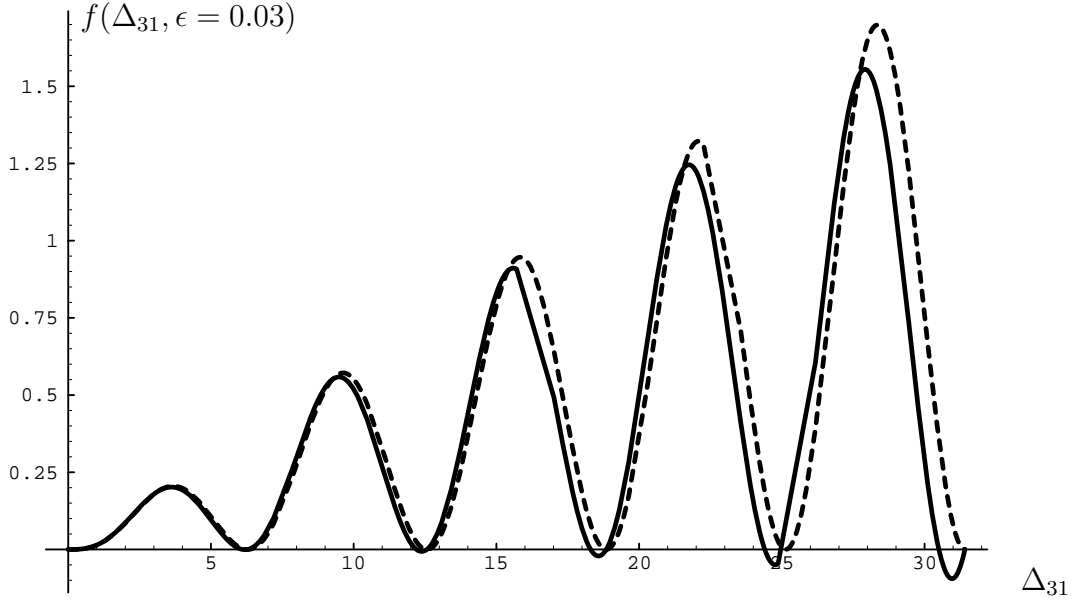}
}
\end{picture}
\caption{Graph of $f(\Delta_{31},\epsilon)$
for $\epsilon = 0.03$.
The solid line and the dashed line represent the exact expression
eq.\ref{eq:f} and the approximated one eq.\ref{eq:fapp}, respectively.
The approximated $f$ has peaks
at $\Delta_{31}=3.67, 9.63,
15.8, \cdots$ irrespectively of $\epsilon$.
}
\label{fig1}
\end{figure}

Fig.\ref{fig1} shows the graph of
$f(\Delta_{31},\epsilon=0.03)$. The approximation eq.\ref{eq:fapp}
works very well up to $|\epsilon \Delta_{31}| \sim 1$.
In the following we will use
eq.\ref{eq:fapp} instead of eq.\ref{eq:f}.
We see many peaks of $f(\Delta_{31},\epsilon)$ in fig.\ref{fig1}.
In practice, however, we do not see such sharp peaks
but observe the value averaged around there, for
$\Delta_{31}$ has a spread due to the energy spread of
neutrino beam ($|\delta\Delta_{31}/\Delta_{31}| = |\delta E/E|$).
In the following we will assume $|\delta\Delta_{31}/\Delta_{31}|
= |\delta E/E| = 20\%$\cite{Nishikawa} as a typical value.


Table \ref{table1} gives values of
$f(\Delta_{31},\epsilon)/\epsilon$ at the first several peaks
and the averaged values around there.

We see the T violation effect,
\begin{eqnarray}
<P(\nu_\alpha \rightarrow \nu_\beta)
- P(\nu_\beta \rightarrow \nu_\alpha)>_{20\%}
= 4 J <f>_{20\%} = J \epsilon \times
\left\{
\begin{array}{c}
25.9 \\
56.0 \\
62.4\\
\vdots
\end{array}
\right.
\hbox{\ for\ }
\Delta_{31} =
\left\{
\begin{array}{c}
3.67 \\
9.63 \\
15.8 \\
\vdots
\end{array}
\right.
\label{eq:atbest2}
\end{eqnarray}
at peaks for neutrino beams with 20 \% of energy spread.
Note that the averaged peak values decrease 
with the spread of neutrino energy.

Which peak we can reach depends on
$\delta m_{31}^2, L$ and $E$. The first peak
$\Delta_{31} = 3.67$ is reached, for example
by $\delta m_{31}^2 = 10^{-2}$ eV$^2$,
$L=250$ km
(for KEK-Kamiokande long base line experiment)
and neutrino energy $E = 1.73$ GeV.
In this case
we see the T(CP) violation effect
at best of $|25.9 J\epsilon| \le
2.50 \epsilon$ since we have a bound on $J$ as eq.\ref{eq:Jmax}.

\begin{table}
\begin{center}
\begin{tabular}{c|c|c|c}
$\Delta_{31}$ & \ $f/\epsilon$\  & $< f/\epsilon>_{10\%}$ &
$<f/\epsilon>_{20\%}$\\
\hline&&&\\
3.67&6.84&6.75&6.48\\
9.63&19.1&17.6&14.0\\
15.8&31.5&25.7&15.6\\
\vdots&\vdots&\vdots&\vdots
\end{tabular}
\end{center}
\caption{The peak values of
$f(\Delta_{31},\epsilon)/\epsilon$ and the corresponding averaged
values. Here
$<f/\epsilon>_{20\%(10\%)}$ is
a value of $f(\Delta_{31},\epsilon)/\epsilon
=\Delta_{31}(1-\cos\Delta_{31})$(see eq.\ref{eq:fapp}) averaged 
over the range $0.8\Delta_{31}\sim 1.2\Delta_{31}$
($0.9\Delta_{31}\sim 1.1\Delta_{31}$).
}
\label{table1}
\end{table}

\vspace*{0.5cm}

\noindent
Situation 2. \ $\Delta_{31} \gg 1$.

Because $\sin\Delta_{32}$ and
$\sin\Delta_{13}$ oscillate rapidly and vanish after
being averaged over the energy spread in this case,
the oscillatory part $f$ is dominated
by $\sin\Delta_{21}$. 
Since $f$ has now a bound $|f| \le 1$ instead of
eq.\ref{eq:fmax},
the T violation effect $4 J f$ is bounded as $|4 J f| \le |4 J|$.
(For energy spread of $10 \sim 20 \%$  of
neutrino beam\cite{Nishikawa},
$\Delta_{31} > 30$ is enough for $\sin\Delta_{32}$ and
$\sin\Delta_{13}$ to oscillate  rapidly and vanish after
being averaged.)

\section{CP violation}

There are a variety of possible combinations of the
parameters, three mixing angles,
two mass differences  and a CP violating phase.
When we consider only the atmospheric 
neutrino anomaly to be attributed to the neutrino oscillation,
we can take the mass differences, $\delta m^2_{21}$ and
$\delta m^2_{31}$(and hence $\delta m^2_{32}$),
to be comparable, while when we consider
both the solar and the atmospheric neutrino anomalies
to be attributed to the neutrino oscillation,
we expect $\delta m^2_{21}$ and
$\delta m^2_{31}$ to be ``disparate'',
$\delta m^2_{21}/\delta m^2_{31} \ll 1$.

We investigate how large the CP violation effect can be
in neutrino oscillation for the above two cases.

\subsection{Comparable mass difference case}

Let's examine the case of
mass differences to be of the same order of magnitude.

We use a parameter set that
$(\delta m^2_{21},\delta m^2_{31}) = (3.8,1.4)
\times 10^{-2}$eV$^2$,
$(\omega,\phi,\psi) =(19^\circ,43^\circ,41^\circ)$
and $\delta$ is arbitrary,
derived
by Yasuda\cite{Yasuda} through the analysis of the
atmospheric neutrino anomaly.
Here the matter effect\cite{Wolf,MS} is negligibly small
and eq.\ref{eq:cpt} is available.

With use of eq.\ref{eq:jf}, eq.\ref{eq:J},
and eq.\ref{eq:cpt} this parameter set gives
the CP violation effect
\begin{eqnarray}
P(\nu_\alpha \rightarrow \nu_\beta) -
 P(\bar\nu_\alpha \rightarrow \bar\nu_\beta)
=  0.22 \sin\delta f(x),
\label{eq:CP1}
\end{eqnarray}
where
\begin{eqnarray}
f(x) = ( \sin 3.8x + \sin 2.4x - \sin 1.4x),
\label{eq:CP1f}
\end{eqnarray}
and
\begin{eqnarray}
 x = 2.5 {(L/100{\rm km}) \over (E/{\rm GeV})}.
\end{eqnarray}

\begin{figure}
\unitlength=1cm
\begin{picture}(16,6)
\unitlength=1mm
\put(135,38){$x$}
\put(0,75){$f(x) \equiv \sin 3.8x + \sin 2.4x - \sin 1.4x$}
\centerline{
\epsfxsize=13cm
\epsfbox{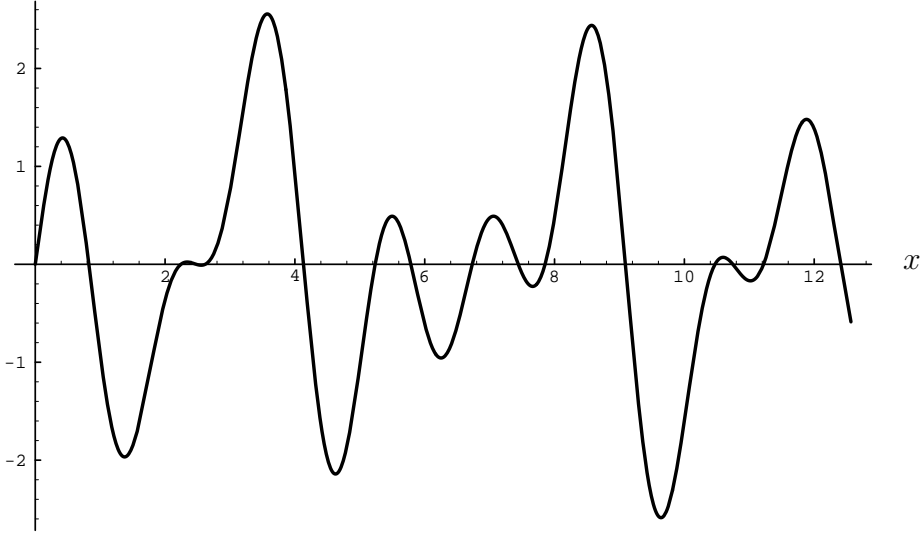}
}
\end{picture}
\caption{Graph of $f(x)$ of eq.\ref{eq:CP1f}.
There are high peaks(positive or negative) at
$x=0.42,1.4,3.6,4.6
\cdots$.
Values of $f(x)$ at peaks averaged over energy
spread of 10\%$\sim$ 20\% are
$<f(0.42)>=1.3\sim 1.3,
<f(1.4)>=-1.9\sim -1.8,
<f(3.6)>=2.2\sim 1.4,
<f(4.6)>=-1.5\sim -0.40\cdots$.}
\label{fig2}
\end{figure}

Fig.\ref{fig2} shows the oscillatory part $f(x)$.
There are many peaks $f(x)$ showing the possibility
to observe the large CP violation effect. For
example, we may see very large difference 
between the transition probabilities,
$<P(\nu_\alpha \rightarrow \nu_\beta) -
P(\bar\nu_\alpha \rightarrow \bar\nu_\beta)>_{20\%}$
$\sim 0.4\sin\delta$ for $L=250$ km(for KEK-Kamiokande experiment)
and $E\sim$4.5 GeV 
corresponding to $x\sim 1.4$, if we have large $\sin\delta$.

Incidentally we may remark that
the survival probability of solar neutrino is calculated
to be 0.45 for those mixing angles.
This value is consistent with both two gallium 
experiments\cite{Ga1,Ga2} and Kamiokande experiment\cite{Kam},
but it is inconsistent with
Homestake result\cite{Cl}, if all the solar neutrino anomaly
should be attributed to the neutrino oscillation\cite{Minakata}.

In conclusion we may see a
large CP violation effect 
when we have comparable mass differences.
In this respect we note that the
long base line experiments are urgently desirable.

\subsection{Disparate mass difference case}

Next we consider the ``disparate'' mass difference case
$\delta m^2_{21}/\delta m^2_{31}\ll1$.

The case $\delta m^2_{31} \sim 1$ eV$^2$ and
$\delta m^2_{21} \sim 10^{-2}$ eV$^2$ is
favoured by the hot dark matter scenario\cite{hdm}
and the atmospheric neutrino anomaly. This case is
already analysed by Tanimoto\cite{Tanimoto}
and we will not discuss it here.

The case
$\delta m^2_{31} \sim 10^{-2}$ eV$^2$ and
$\delta m^2_{21} \sim 10^{-4}$ eV$^2$ could typically explain
the anomalies of the atmospheric and the solar neutrinos\cite{FLM}.
In this case we cannot neglect
the matter effect\cite{Wolf,MS}
\begin{equation}
2\sqrt{2}G_F n_e E \sim 2\times10^{-4} {\rm eV}^2
\left({E \over {\rm GeV}}\right)
\left({n\over 3{\rm g/c.c}}\right),
\label{eq:mattereffect}
\end{equation}
where $n_e$ is the electron number density of the
earth and $n$ is the matter density of the surface of the earth,
since it is greater than $\delta m^2_{21}$.
It requires to subtract such effect in order to
deduce the pure CP violation effect\cite{prep}.
In principle it is possible,
because the matter effect is proportional to $E$
while $\delta m^2_{21}$ is constant.

\section{T violation}

In the matter with  constant density\footnote{
Note that the time reversal of $\nu_\alpha \rightarrow
\nu_\beta$ requires the exchange of the production point and the
detection point and the time reversal of
$P(\nu_\alpha \rightarrow \nu_\beta)$
in matter is in general different from
$P(\nu_\beta \rightarrow\nu_\alpha)$\cite{KP}.},
 we have a pure
T violation effect $P(\nu_\alpha \rightarrow
\nu_\beta) - P(\nu_\beta \rightarrow
\nu_\alpha)$, though we do not observe a pure CP violation effect
because of an apparent CP violation due to matter.

\subsection{T violation in matter}
When a neutrino is
in matter, its matrix of effective mass squared
$M_m^2$ of weak eigenstates is\cite{KP,Toshev}
\begin{eqnarray}
M_m^2 &=& U \pmatrix{0&&\cr&\delta m_{21}^2&\cr&&\delta m_{31}^2}
U^\dagger + \pmatrix{a&&\cr&0&\cr&&0},
\end{eqnarray}
where $a = 2\sqrt{2} G_F n_e E$ and $U$ is given by
eq.\ref{eq:mixingmatrix}.
This is diagonalized by a
mixing matrix $U_m$ as $M_m^2 = U_m
{\rm diag}(\tilde m_1^2,\tilde m_2^2, \tilde m_3^2)
U_m^\dagger$. It
is written with a real unitary
(orthogonal) matrix $\tilde U$ as
\begin{equation}
U_m = \exp (i\psi\lambda_7) \Gamma \tilde U.
\end{equation}

With arguments analogous to \S 2.1
we have the T violation effect,
\begin{eqnarray}
P(\nu_\alpha \rightarrow
\nu_\beta) - P(\nu_\beta \rightarrow
\nu_\alpha) = 4 J_m f_m,
\end{eqnarray}
where
\begin{eqnarray}
J_m 
&=& -{\rm Im}U_{m\beta 1}U^*_{m\beta 2}U^*_{m\alpha 1}U_{m\alpha 2}
\nonumber\\
&=& \sin\psi \cos\psi \tilde U_{11}\tilde U_{12}\tilde U_{13}
\sin\delta,\label{eq:Jm}\\
f_m &=& \sin{\tilde m^2_2-\tilde m^2_1\over 2E}L
+\sin{\tilde m^2_3-\tilde m^2_2\over 2E}L
+\sin{\tilde m^2_1-\tilde m^2_3\over 2E}L.
\end{eqnarray}

We get
\begin{eqnarray}
D_m &\equiv& {\rm diag}(\tilde m_1^2,
\tilde m_2^2, \tilde m_3^2)
\nonumber\\
&=& U_m^\dagger M_m^2 U_m
\nonumber\\
&=& \tilde U^\dagger\left\{U_\phi U_\omega
\pmatrix{0&&\cr&\delta m_{21}^2&\cr&&\delta m_{31}^2}
U_\omega^T U_\phi^T
+ \pmatrix{a&&\cr&0&\cr&&0}
\right\}\tilde U
\nonumber\\
&=&\tilde U^\dagger\left\{
\pmatrix{a+\delta m_{31}^2 \sin^2\phi &0&\delta m_{31}^2
\cos\phi\sin\phi\cr
0&0&0\cr
\delta m_{31}^2\cos\phi\sin\phi&0&\delta m_{31}^2\cos^2\phi} 
+ \delta m_{21}^2 U_\phi U_\omega \pmatrix{0&&\cr &1&\cr&&0}
U_\omega^T U_\phi^T
\right\}\tilde U,
\nonumber\\
\label{eq:matter}
\end{eqnarray}
where $U_\phi = \exp (i\phi\lambda_5)$ and
$U_\omega = \exp (i\omega\lambda_2)$.

An exact result for $U_m$ and $D_m$ is given in \cite{ZS}, though
their result is rather complicated. Here we show
a simple expression for $U_m$ and $D_m$ in
the case $\delta m_{21}^2 \ll a,
\delta m_{31}^2$. We derive $U_m$ and $D_m$ in this case
using perturbation with respect to small $\delta m_{21}^2$.

{}First we decompose $\tilde U = U_0 V$ and diagonalize
by $U_0$ the first term of the parenthesis \{\}
of eq.\ref{eq:matter}, the
eigenvalues of which we denote by $\Lambda_i$'s. We find
\begin{eqnarray}
U_0 = \exp (i\phi'\lambda_5) \hbox{\ with\ }
\tan 2\phi'= {
 \delta m_{31}^2 \sin 2\phi
\over \delta m_{31}^2\cos 2\phi -a},
\end{eqnarray}
and
\begin{eqnarray}
\left\{
\begin{array}{lcl}
\Lambda_1 &=& {(a+\delta m^2_{31}) - \sqrt{(a+\delta m^2_{31})^2-
4a\delta m^2_{31}\cos^2\phi} \over 2},\\
\Lambda_2 &=& 0,\\
\Lambda_3 &=& {(a+\delta m^2_{31}) + \sqrt{(a+\delta m^2_{31})^2-
4a\delta m^2_{31}\cos^2\phi} \over 2}.
\end{array}
\right.
\end{eqnarray}

We have
\begin{eqnarray}
D_m &=&
V^\dagger\left\{\pmatrix{\Lambda_1&&\cr&\Lambda_2&\cr&&\Lambda_3}
+ \delta m_{21}^2 U_{\phi-\phi'} U_\omega \pmatrix{0&&
\cr&1&\cr&&0} U_\omega^T U_{\phi-\phi'}^T
\right\}V
\nonumber \\
&\equiv&V^\dagger\left\{{\rm diag}(\Lambda_1,\Lambda_2,\Lambda_3)
+ \delta m_{21}^2 H
\right\}V.
\end{eqnarray}

Next we diagonalize the whole $M_m^2$ by $V$ with
perturbation with respect to small $\delta m^2_{21}$.

At the zeroth order of $\delta m_{21}^2$, we have
$\tilde m^2_i = \Lambda_i,$
$V_{ij} = \delta_{ij}$, and $\tilde U= U_0$ which gives
$\tilde U_{12} = (U_0)_{12}
=0$ and hence $J_m=0$(see eq.\ref{eq:Jm}).

At the first order of perturbation, we have 
\begin{eqnarray}
\tilde m^2_i &=&  \Lambda_i + \delta m_{21}^2 H_{ii},\\
V_{ij}&=& 
\left\{
\begin{array}{lll}
1&\hbox{for} &i=j\\
\delta m_{21}^2 {H_{ij} \over \Lambda_j - \Lambda_i}
&\hbox{for} &i\ne j
\end{array}
\right. ,
\end{eqnarray}
and with eq.\ref{eq:Jm}
\begin{equation}
J_m = - {\delta m_{21}^2 \over a} {\delta m_{31}^2
\over \{(\delta m_{31}^2 + a)^2 - 4\delta m_{31}^2 a
\cos^2\phi\}^{1/2}}
\sin\omega\cos\omega\sin\psi\cos\psi\sin\phi\sin\delta.
\label{eq:Jmp}
\end{equation}

\subsection{Most likely case:
$\delta m^2_{21} \ll a \ll \delta m^2_{31}$}

It seems most likely to be realized that
$\delta m^2_{21} \ll a \ll \delta m^2_{31}$ as is discussed
in \S 3.2. Here we study this case in detail.
Since $J_m$ is $O(\delta m^2_{21})$ we neglect
$O(\delta m^2_{21})$ in estimating $f_m$.
We also
neglect $O(a^2)$ since $a/\delta m_{31}^2\ll 1$.

Then we have the effective masses
\begin{eqnarray}
\tilde m^2_1 &\simeq& \Lambda_1 \simeq a\cos^2\phi ,
\nonumber\\
\tilde m^2_2&\simeq&  \Lambda_2 \simeq 0 ,\\
\tilde m^2_3&\simeq&  \Lambda_3 \simeq 
\delta m_{31}^2 + a\sin^2\phi.
\nonumber
\end{eqnarray}
and ``mass difference ratio''
\begin{eqnarray}
\epsilon_m = 
{\tilde m^2_2 - \tilde m^2_1 \over \tilde m^2_3 - \tilde m^2_2}
\simeq - { a\cos^2\phi \over \delta m_{31}^2}.
\end{eqnarray}
Note that $|\epsilon_m| \ll 1$.

We find
\begin{eqnarray}
J_m \sim  - {\delta m_{21}^2 \over a}
\sin\omega\cos\omega\sin\psi\cos\psi\sin\phi\sin\delta,
\end{eqnarray}
and 
\begin{eqnarray}
J_m \epsilon_m = J \epsilon.
\end{eqnarray}

Using the argument similar to that used to derive
eq.\ref{eq:atbest2}, we obtain the T violation effect
\begin{eqnarray}
<P(\nu_\alpha \rightarrow \nu_\beta)
- P(\nu_\beta \rightarrow \nu_\alpha)>_{20\%}
= J_m \epsilon_m \times
\left\{
\begin{array}{c}
25.9 \\
56.0 \\
62.4\\
\vdots
\end{array}
\right.
= J \epsilon \times
\left\{
\begin{array}{c}
25.9 \\
56.0 \\
62.4\\
\vdots
\end{array}
\right.,
\end{eqnarray}
at peaks, where
we choose the mean neutrino energy $E$ to satisfy
(see Table \ref{table1})
\begin{eqnarray}
\Delta_{31} = \delta m^2_{31} {L \over 2E} = 3.67, 9.63, 15.8
\ldots
\end{eqnarray}

According to the analysis 
by Fogli {\it et al}\cite{FLM}, $J/\sin\delta$
$\sim0.06$ and
 $\epsilon \sim 10^{-2}$ are allowed\footnote{
Here $\sin\omega \sim 1/2, \sin\psi\sim1/\sqrt{2}$
and $\sin\phi
= \sqrt{0.1}$.} 
 for example.
Then
\begin{equation}
<P(\nu_\alpha \rightarrow \nu_\beta) - P(\nu_\beta \rightarrow
\nu_\alpha)>_{20\%}
= \left({J/\sin\delta\over 0.06}\right)
\left({\epsilon\over 10^{-2}}\right) \sin\delta \times
\left\{
\begin{array}{c}
0.015 \\
0.033 \\
0.037\\
\vdots
\end{array}
\right. .
\end{equation}

\section{Summary}

We have examined the CP and T violation in the neutrino
oscillation, and analysed how large the violation can be by taking
account of the constraints of the neutrino experiments. 

In case of the comparable mass differences of $\delta m^2_{21},
\delta m^2_{31}$ and  $\delta m^2_{32}$
in the range $10^{-3}$ to $10^{-2}$ eV$^2$, 
which is consistent with the analysis of the atmospheric
neutrino anomalies, it is found that there is a possibility
that the CP violation effect is large enough to be observed
by 100 $\sim$ 1000 km base line experiments if the CP
violating parameter sin$\delta$ is sufficiently large. 

In case that $\delta m^2_{21}$
is much smaller than the matter parameter 
``$a$'' and $\delta m^2_{31}$,
which is favoured both by the solar and atmospheric
neutrino anomalies, we have derived a simple formula for the T 
violation effect. We note that the probability of CP or T violation
effect
should vanish for  $\delta m^2_{21} \rightarrow$ 0, 
and therefore be proportional to  
$\delta m^2_{21}/\delta m^2_{31}$, $\delta m^2_{21}/(E/L)$
or $\delta m^2_{21}/a$ by the dimensional analysis. Our 
calculation confirms this expectation. If the solar and
atmospheric neutrino anomalies are both attributed to
the neutrino oscillation, 
the CP violation test is found difficult since
matter effect is larger than the pure CP violation effect.
How to extract the matter 
effect in such a case will be discussed in a separate
paper\cite{prep}. 

In conclusion the long base line neutrino oscillation
experiments are very important and desirable
to study not only neutrino masses and 
mixings but the CP or T violation in the lepton 
sector and there is some 
possibility to find such effect explicitly.

We finally express our 
thanks to Prof. K. Nishikawa for valuable discussions and 
communications.

\end{document}